\documentclass[floatfix,pra,10pt,twocolumn,superscriptaddress, noeprint,nofootinbib]{revtex4-2}
\usepackage{amsmath,bm}
\usepackage{ulem}
\usepackage{latexsym}
\usepackage{amssymb}
\usepackage{graphics,epstopdf}
\usepackage{hyperref}
\usepackage{xcolor}

\newcommand{\splitatcommas}[1]{%
  \begingroup
  \begingroup\lccode`~=`, \lowercase{\endgroup
    \edef~{\mathchar\the\mathcode`, \penalty0 \noexpand\hspace{0pt plus 1em}}%
  }\mathcode`,="8000 #1%
  \endgroup
}

\usepackage{subcaption}
\usepackage[percent]{overpic}
\usepackage{comment}
\usepackage{mathtools}
\usepackage{braket}

\newcommand{\mr}[1]{\mathrm{#1}}

\MakeRobust{\eqref}

\usepackage{subcaption}

\usepackage{graphicx}
\usepackage{float}
\usepackage{amssymb}

\usepackage{caption}
\usepackage{subcaption}

\usepackage{etoolbox}
\AtBeginEnvironment{thebibliography}{\normalem}

\begin{document}
\title{Asymptotic freedom in the 
dephased charging of quantum batteries} 
\author{Chayan Purkait}
\email{0624394@zju.edu.cn}
\thanks{Contributed equally and may be contacted.}
\affiliation{Department of Physics and Zhejiang Institute of Modern Physics, Zhejiang University, Hangzhou, Zhejiang 310027, China}
\author{B. Prasanna Venkatesh}
\email{prasanna.b@iitgn.ac.in}
\thanks{Contributed equally and may be contacted.}
\affiliation{Indian Institute of Technology Gandhinagar, Palaj, Gujarat 382055, India}
\author{Gentaro Watanabe}
\email{gentaro@zju.edu.cn}
\thanks{Corresponding author; contributed equally.}
\affiliation{Department of Physics and Zhejiang Institute of Modern Physics, Zhejiang University, Hangzhou, Zhejiang 310027, China}
\affiliation{Zhejiang Province Key Laboratory of Quantum Technology and Device, Zhejiang University, Hangzhou, Zhejiang 310027, China}

\begin{abstract}
Quantum batteries—small-scale energy storage devices based on quantum systems—offer the potential for enhanced charging performance through quantum effects such as coherence and collectivity. In this work, we study the collective charging of quantum batteries 
consisting of \( N \) qubits, coupled to a driven qubit charger in a star configuration, with controlled pure dephasing acting on the charger. We investigate how an ``asymptotic freedom''-like behavior—in which all the energy deposited into the battery can be extracted as work, resulting in the ergotropy-to-energy ratio approaching unity—can emerge in the steady state of the battery. We show that the ergotropy-to-energy ratio increases with the number of qubits and approaches unity asymptotically as $1-O(1/N)$. In the large-$N$ limit, the emergence of approximate ground-state degeneracy of the collective battery system leads to this asymptotic freedom behavior, despite the battery state remaining mixed. We also discuss the scaling behavior of the charging time of the battery with $N$.
\end{abstract}

\maketitle

\section{\label{sec:Introduction} Introduction}

The ability to control, manipulate, and measure nanoscale quantum systems has motivated the study of microscopic analogs of macroscopic thermodynamic devices, such as batteries, thermal machines, and thermoelectric circuits \cite{Quach2023,Bhattacharjee2021,myers2022quantum}. 
Ideas from quantum thermodynamics \cite{Vinjanampathy01102016} play a crucial role in the design of these devices as well as in developing strategies to optimize their performance. A central goal of such studies is to identify quantum phenomena that offer advantages with no classical counterpart. Storing energy and powering devices at the nanoscale lie at the heart of many quantum technologies, where quantum batteries (QBs) can play a pivotal role. A substantial body of research has been devoted to characterizing the optimal performance of QBs in terms of total stored energy, ergotropy (the extractable part of the energy by some unitary operations), and charging power \cite{Bhattacharjee2021,Campaioli2018,Andolina2018,GarciaPintos2020,Chen2022,Dou2022,PhysRevA.109.022607,RevModPhys.96.031001,Alicki2013,Hovhannisyan2013,Gallacher2015,Campaioli2017,Ferraro2018,Andolina2019,Ito2020,Rossini2020,Julia-Farre2020, Watanabe2020,Gyhm2022,Yan2023,Yang2023a,Gyhm2024,SongPRL2024,SongPRL2025,BaiPRA2020}.

Early studies modeled quantum batteries as closed systems, with their charging processes actuated by unitary transformations. Since no quantum system is closed in the true sense, dissipative charging of quantum batteries has gained a lot of interest in recent times with studies considering various realistic scenarios including the effect of environment \cite{Liu2019,Pirmoradian2019,Barra2019,Farina2019,Quach2020,Tabesh2020,Kamin2020,Xu2021,Santos2021,Ghosh2021,Quach2022,Mayo2022,rodriguez2024optimal,Yang2023b,Rodriguez2023,Dou2023,PhysRevLett.132.210402,gangwar2024coherently,PhysRevA.110.032211,PhysRevLett.134.130401,PhysRevLett.132.210402}.
Moreover, quantum control based schemes to stabilize the stored energy and ergotropy \cite{Mitchison2021,Yao2021,Yao2022,Gherardini2020,Malavazi2025}, exploiting collective dissipation to protect \cite{Quach2020,Xu2024} or enhance charging \cite{PhysRevLett.134.130401}, and exploiting reservoir engineering to create non-reciprocity to design better batteries \cite{PhysRevLett.132.210402} have been proposed.

Recently, a dephasing-enabled stable charging scheme for quantum batteries has been proposed by us \cite{shastri2025dephasing}. It has been demonstrated that optimal pure dephasing on a driven quantum charger can lead to the fast charging of quantum batteries. However, in open-system setups, the quality of charging is generally poor, as the ergotropy is often significantly less than the total stored energy due to decoherence. This raises a natural question: how can we enhance the quality of charging such that most of the stored energy is extractable as work, while the locked energy remains negligible?

The collective charging of a quantum battery, in which multiple battery elements are charged simultaneously, has attracted considerable attention in recent studies \cite{Andolina2019,PhysRevA.111.022222,Julia-Farre2020,Rossini2020,Quach2022,Ferraro2018,10.3389/fphy.2022.1097564,doi:10.7566/JPSJ.91.124002,Le2018,Gallacher2015,Campaioli2017,Ito2020,PhysRevA.110.032211,PhysRevA.110.032211,Carrasco2022,PhysRevLett.134.130401}. The main goal of these studies is to investigate how higher performance can be achieved in the collective setup compared to its parallel counterpart. The studies demonstrated that the quality of charging can be improved in the collective setup: in the limit of a large number $N$ of batteries, the locked energy (non-extractable energy) becomes negligible with respect to the total stored energy. All the deposited energy can then be extracted as work, resulting in the ergotropy-to-charge ratio approaching one. This phenomenon is known as asymptotic freedom \cite{Andolina2019,PhysRevA.111.022222}. On the other hand, superextensive charging power, where the average charging power scales faster than linearly with the number of batteries ($\bar{P} \propto N^{\alpha>1}$), can be obtained in the collective setup. This is enabled by quantum entanglement, in particular the trajectory length in Hilbert space can be reduced by passing through highly entangled states during charging operations \cite{Julia-Farre2020,Rossini2020,Quach2020,Gallacher2015,Campaioli2017}.
Furthermore, the cooperative effect that enhances the effective quantum coupling between the energy source and the battery can also give rise to superextensive charging power \cite{Quach2022,Ferraro2018,10.3389/fphy.2022.1097564,doi:10.7566/JPSJ.91.124002,Le2018,PhysRevA.111.022222}.

In this work, we explore how the quality of charging can be improved in the collective charging setup under dephasing. In previous studies, asymptotic freedom has been shown to emerge in the transient regime of quantum battery charging~\cite{Andolina2019,PhysRevA.111.022222}. In~\cite{Andolina2019}, the authors analyzed fully unitary dynamics in the closed Tavis–Cummings model and identified the optimal charging time as the first maximum of ergotropy. Despite the high total energy stored during charging, correlations between the charger and the battery can suppress the amount of extractable work. Remarkably, they demonstrated that, in the thermodynamic limit, the battery exhibits asymptotic freedom, whereby the fraction of locked (non-extractable) energy vanishes as $N \to \infty$. Building on this foundation,~\cite{PhysRevA.111.022222} extended the analysis to open Dicke and Tavis–Cummings batteries subject to single-atom dissipation and dephasing, showing that genuine charging persists only during transient dynamics. This is because, over long periods, evolution is governed by the environment rather than the charger, and the system relaxes to a unique steady state that is independent of the charger's initial state: irrespective of the initial amount of energy in the charger, the energy in the steady state is the same. This makes steady-state charging not meaningful in their setup. Evaluating performance at the time of maximum ergotropy, they confirmed that asymptotic freedom survives even in the presence of local noise. 
We explore how the asymptotic freedom behavior can be achieved in the steady-state dephased charging of quantum batteries. 
\begin{figure}
    \includegraphics[width=0.7 \linewidth]{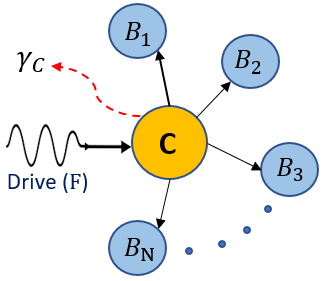}
    \caption{Schematic diagram of a quantum battery $\mathrm{B}$ consisting of $N$ independent identical systems, coupled to a quantum charger $\mathrm{C}$ driven with strength $F$, where the charger is also subject to dephasing at a rate $\gamma_{\mathrm{C}}$.
    }
 \label{fig:schematic}
\end{figure}

We consider the charging of quantum batteries consisting of \( N \) identical systems coupled to a driven charger system, in a star configuration, as shown schematically in Fig.~\ref{fig:schematic}. In addition, a controlled pure dephasing process also acts on the charger. Note that we differ from the previous studies \cite{Andolina2019,PhysRevA.111.022222} because of the driving and dephasing of the charger. While the choice of the star-shaped charger–battery configuration is motivated by recent experimental studies of quantum batteries on NMR platforms \cite{Joshi2022,PandePRA2017,LiuQST2024,AlgabaPRR2022}, pure dephasing is an important and ubiquitous channel of decoherence in open quantum systems. In particular, pure dephasing typically arises from the coupling of a system operator commuting with the system Hamiltonian to a noisy environment. 
Under such a setting, we obtain the steady-state charging of the batteries. Additionally, we do not need any fine tuning of the initial state to establish our main results. In particular, we can start with the ground state of the charger and battery. Given the setup, the goal of our work is to investigate how the relevant figures of merit for battery operation scale with the number of individual battery systems $N$. Choosing the battery and charger system as qubits for concreteness, we show that the ergotropy-to-energy ratio increases with the number of qubits. Remarkably, we find that in the large-$N$ limit, the emergence of approximate ground-state degeneracy causes the ergotropy-to-energy ratio to asymptotically approach unity—referred to as ``asymptotic freedom''—despite the battery state remaining mixed. Therefore, we achieve ``asymptotic freedom'' in the steady state. This is the central result of our manuscript. Furthermore, in agreement with previous work with single qubit battery set-ups \cite{shastri2025dephasing}, we obtain an optimal dephasing rate, $\gamma_{\mathrm{C}}^*$, of the charger that enables fast charging. Interestingly, we find that the optimal charging time $\tau^* \propto 1/\gamma_{\mathrm{C}}^*$ exhibits distinct scaling behaviors with the number of qubits $N$, namely $\tau^* \sim N^b$ with the value of $b$ dependent on the strength of the charger driving.

The paper is organized as follows. Section~\ref{sec:Setup} introduces the charger–battery model and the figures of merit, while Section~\ref{sec:Results} presents the main results along with a detailed analysis. Finally, Section~\ref{sec:Conclusion} summarizes the key findings and conclusions.


\section{\label{sec:Setup} Setup and figures of merit}

\begin{figure}
    \centering
    \includegraphics[width=0.7 \linewidth]{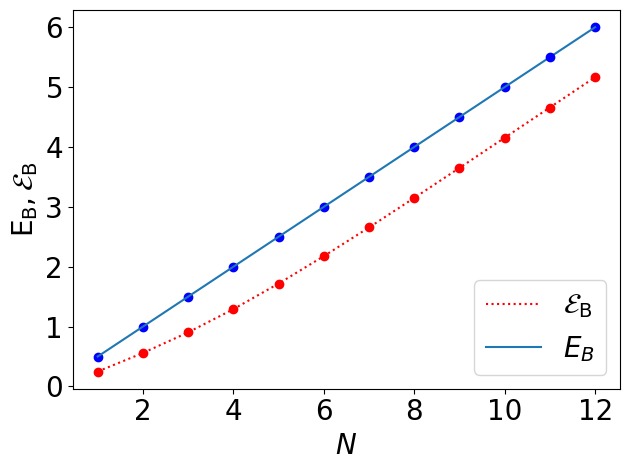}
    \caption{Variation of the total energy $E_{\mathrm{B}}$ and ergotropy $\mathcal{E}_{\mathrm{B}}$ of the battery in the steady state under intermediate driving ($F/g = 0.5$) as a function of the number of qubits $N$. The other parameters are $\omega_{\mathrm{B}} = \omega_{\mathrm{C}} = \omega_{\mathrm{d}} = 1$ and $g = 1.0\,\omega_{\mathrm{B}}$.}
 \label{fig:energy and ergotropy for F = 0.5}
\end{figure}

\begin{figure*}
    \centering
    \includegraphics[width=0.93 \linewidth]{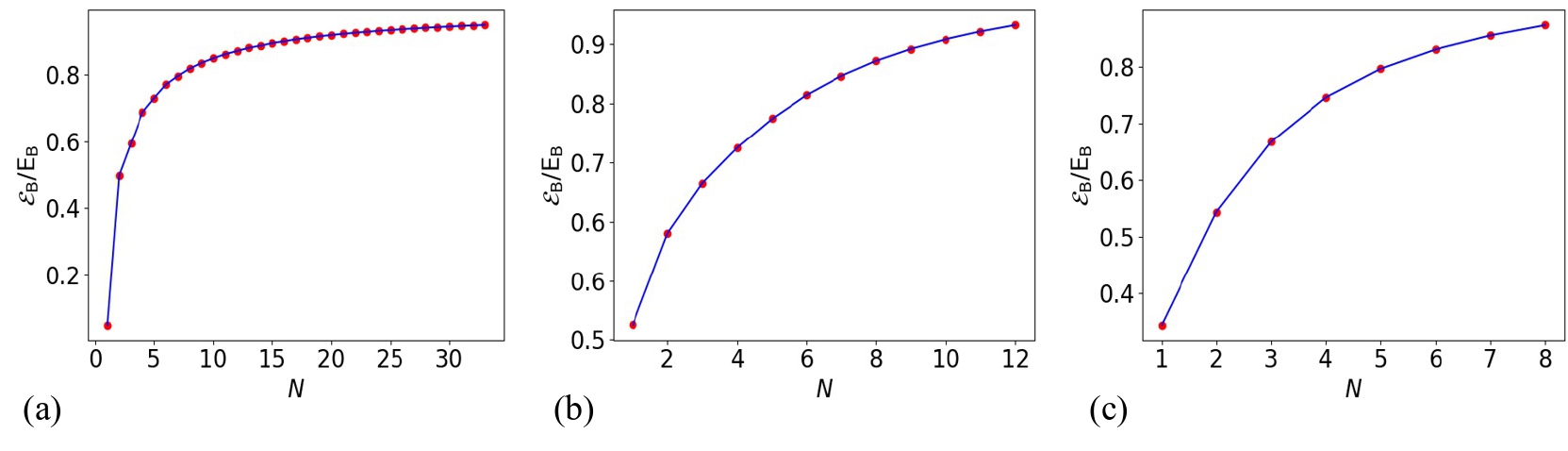}
    \caption{Variation of steady-state ergotropy relative to the total energy ($\mathcal{E}_{\mathrm{B}}/E_{\mathrm{B}}$) as a function of the number of qubits ($N$) for (a) strong driving $F = 10 \omega_{\mathrm{B}}$ $(F/g = 10)$, (b) intermediate driving $F = 0.5 \omega_{\mathrm{B}}$ $(F/g = 0.5)$, and (c) weak driving $F = 0.2 \omega_{\mathrm{B}}$ $(F/g = 0.2)$. The other parameters are the same as in Fig.~\ref{fig:energy and ergotropy for F = 0.5}. The ratio $\mathcal{E}_{\mathrm{B}}/E_{\mathrm{B}}$ is given by $\mathcal{E}_{\mathrm{B}}/E_{\mathrm{B}} = 1-2\delta(N) N^{-1}$, where $\delta(N)$ is the population of the excited states of the passive state at a given $N$ (see Appendix~\ref{app:A} for a detailed derivation of this relation).}
    \label{fig:Steady-state ergotropy relative to the total energy}
\end{figure*}
The Hamiltonian for a quantum battery system $\mathrm{B}$ consisting of $N$ qubits coupled to a common charger system $\mathrm{C}$ in a star-like configuration (see Fig.~\ref{fig:schematic}) is given by (we take $\hbar=1$ throughout)
\begin{equation}
\begin{aligned}
\hat{H}(t)=&~\hat{H}_{\mathrm{C}}+\hat{H}_{\mathrm{B}}+\hat{H}_{\mathrm{CB}}+\hat{H}_{\mathrm{d}}(t)\\= &~\omega_{\mathrm{C}} \hat{\sigma}_{\mathrm{C}}^{+} \hat{\sigma}_{\mathrm{C}}^{-}+\omega_{\mathrm{B}} \sum_{j=1}^N \hat{\sigma}_{\mathrm{B}, j}^{+} \hat{\sigma}_{\mathrm{B}, j}^{-} \\
&+ \frac{g}{\sqrt N} \left(\hat{\sigma}_{\mathrm{C}}^{+} \sum_{j=1}^N \hat{\sigma}_{\mathrm{B}, j}^{-} + h.c. \right)  \\
&+ F\left(\hat{\sigma}_{\mathrm{C}}^{-} e^{i \omega_{\mathrm{d}} t}+\hat{\sigma}_{\mathrm{C}}^{+} e^{-i \omega_{\mathrm{d}} t}\right).
\end{aligned}
\label{eq:Hamiltonian}
\end{equation}
Here, $\hat{H}_{\mathrm{C}}$ and $\hat{H}_{\mathrm{B}}$ are the bare Hamiltonians of the charger and the battery, respectively; $\hat{H}_{\mathrm{CB}}$ gives the coupling Hamiltonian between the charger and all battery elements that enable charging; $\hat{H}_{\mathrm{d}}$ is the operator representing the coherent driving of the charger that acts as a source of energy. Also, $\hat{\sigma}^{-} (\hat{\sigma}^{+})$ denotes the Pauli lowering (raising) operator, while $\omega_{\mathrm{B}}\left(\omega_{\mathrm{C}}\right)$ represents the frequency of the battery (charger), and $F$ and $\omega_{\mathrm{d}}$ correspond to the drive strength and frequency, respectively. In our study, we primarily consider the resonant case given by $\omega_{\mathrm{B}}=\omega_{\mathrm{C}}=\omega_{\mathrm{d}}$. In the absence of driving ($\hat{H}_{\mathrm{d}} = 0$) and $\omega_{\mathrm{B}}=\omega_{\mathrm{C}}$, the coupling Hamiltonian $\hat{H}_{\mathrm{CB}}$ commutes with $\hat{H}$, ensuring that there is no energetic cost associated with switching the interaction on and off. If $\hat{H}_{\mathrm{d}}$ is nonzero, $\hat{H}_{\mathrm{CB}}$ does not commute with $\hat{H}$. To keep the ratio $F/g$ fixed as $N$ increases, the coupling constant $g$ is scaled by $1/\sqrt{N}$, ensuring that the total coupling strength between the charger and the battery remains finite. This scaling normalizes the collective interaction, ensuring well-behaved dynamics in the thermodynamic limit \cite{kac1963JMP,kastner2025Axriv}.

In addition, the charger system is subject to pure dephasing, modeled using a Gorini-Kossakowski-Sudarshan-Lindblad (GKLS) master equation \cite{gorini1976completely,lindblad1976generators} with a Hermitian jump operator $\hat{L}_{\mathrm{C}}$ satisfying $\left[\hat{L}_{\mathrm{C}}, \hat{H}_{\mathrm{C}}\right]=0$. Consequently, the time evolution of the density matrix $\hat{\rho}$, describing the joint state of the charger and battery systems, is given by 
\begin{equation}\label{maste eq}
  \frac{d }{dt}\hat{\rho}(t) =-i[\hat{H}, \hat{\rho}(t)]+\frac{\gamma_{\mathrm{C}}}{2}\left(2\hat{L}_{\mathrm{C}} \hat{\rho}(t) \hat{L}_{\mathrm{C}}-\left\{\hat{L}_{\mathrm{C}}^2, \hat{\rho}(t)\right\}\right).
\end{equation}
Here, $\gamma_{\mathrm{C}}$ denotes the dephasing rate, and $\{\cdot, \cdot\}$ represents the anticommutator. Choosing the jump operator as $\hat{L}_{\mathrm{C}} \propto \hat{H}_{\mathrm{C}}$ allows the pure dephasing process described in Eq.~\eqref{maste eq} to be interpreted as the result of a continuous weak measurement of the charger's energy. In our study,  the jump operator is taken to be $\hat{L}_{\mathrm{C}}=\hat{\sigma}_{\mathrm{C}}^{+} \hat{\sigma}_{\mathrm{C}}^{-}$.

We assume that all the battery elements and the charger are initially prepared in their respective uncoupled ground states, 
\begin{align}
\hat{\rho}(0) = \ket{0}_{\mathrm{C}\, \mathrm{C}}\!\bra{0} \otimes 
\left(\ket{0}_{\mathrm{B}\, \mathrm{B}}\!\bra{0}\right)^{\otimes N},    \label{eq:initstate}
\end{align}
where $\ket{0}_{\mathrm{C}}$ and $\ket{0}_{\mathrm{B}}$ denote the uncoupled ground states of the charger and each battery element, respectively. 
The evolution governed by Eq.~\eqref{maste eq} leads to an increase in both the total energy and the ergotropy of the battery. The total energy of the battery is defined as
\begin{align}
   E_{\mathrm{B}} &= \operatorname{Tr}_{\mathrm{B}}\!\left[\hat{\rho}_{\mathrm{B}} \hat{H}_{\mathrm{B}}\right],
\end{align}
where $\hat{\rho}_{\mathrm{B}} \equiv \operatorname{Tr}_{\mathrm{C}}[\hat{\rho}]$ is the reduced density matrix of the battery.

Ergotropy \cite{AllahverdyanEPL2004} refers to the maximum extractable work from a quantum system via a cyclic unitary operation. This is determined by finding the minimum possible internal energy of the system’s final state
\begin{equation}\label{erg}
    \mathcal{E}_{\mathrm{B}} = 
E_{\mathrm{B}} - 
\min_{\hat{U}_{\mathrm{B}}} 
\operatorname{Tr}_{\mathrm{B}}\!\left[
\hat{U}_{\mathrm{B}} \hat{\rho}_{\mathrm{B}} 
\hat{U}_{\mathrm{B}}^{\dagger} \hat{H}_{\mathrm{B}}
\right],
\end{equation}
where minimization 
$\min_{\hat{U}_{\mathrm{B}}}$ is taken over all possible unitaries acting in the battery's $2^N$-dimensional Hilbert space. 
The spectral decompositions of the battery state and its Hamiltonian are
\begin{equation}\label{Spectal}
\begin{aligned}
    \hat{\rho}_{\mathrm{B}} = \sum_i \eta_i^{\downarrow}\, \ket{i^{\downarrow}}\!\bra{i^{\downarrow}},\\
\hat{H}_{\mathrm{B}} = \sum_i \varepsilon_i^{\uparrow}\, \ket{\varepsilon_i^{\uparrow}}\!\bra{\varepsilon_i^{\uparrow}},
\end{aligned}
\end{equation}
where the eigenvalues are ordered such that 
$\eta_i^{\downarrow} \ge \eta_{i+1}^{\downarrow}$ and $\varepsilon_i^{\uparrow} \le \varepsilon_{i+1}^{\uparrow}$.
The state that attains this minimum energy takes the form $
\hat{\rho}_{\mathrm{B}}^{\downarrow}=\sum_i \eta_i^{\downarrow}\, \ket{\varepsilon_i^{\uparrow}}\!\bra{\varepsilon_i^{\uparrow}}.
$
By substituting $\hat{\rho}_{\mathrm{B}}^{\downarrow}$ into Eq.~(\ref{erg}), we get 
\begin{align}
    \mathcal{E}_{\mathrm{B}}=\sum_{j, l} \eta_j^{\downarrow} \varepsilon_l^{\downarrow}\left(\left|\left\langle i_j^{\uparrow} \mid \varepsilon_l^{\uparrow}\right\rangle\right|^2-\delta_{j l}\right).
\end{align}
Note that the dynamics given by Eq.~\eqref{maste eq}, with the initial conditions Eq.~\eqref{eq:initstate}, generate permutation invariant Dicke states of maximal angular momentum of the collective battery system, which form a $(N+1)$-dimensional sub-space of the battery's full Hilbert space. Nonetheless, we choose to optimize over the full Hilbert space of the battery in the ergotropy calculation (see Eq.~\eqref{erg}) to ensure that the resulting quantity is the true upper limit of possible work extraction from the given battery state.

In addition to energy and ergotropy, which serve as standard figures of merit, another important parameter characterizing the charging performance of a quantum battery is the charging time $\tau$. We define $\tau$ as the characteristic timescale over which the battery approaches its maximum (steady-state) energy. Operationally, $\tau$ is extracted by fitting the time-coarse-grained energy dynamics to an exponentially damped function,  
\begin{equation}
E_{\mathrm{B}}(t) \approx E_\mathrm{B}(\infty) \Big[ 1 - e^{-t/\tau} \cos(\omega t + \phi) \Big],
\end{equation}
where $E_\mathrm{B}(\infty)$ denotes the saturation energy, $\omega$ is the oscillation frequency, and $\phi$ is a phase offset. Physically, $\tau$ represents the dominant decay timescale of the transient dynamics, quantifying how quickly the system relaxes toward its steady-state energy. A smaller $\tau$ corresponds to faster charging, whereas a larger $\tau$ indicates slower charging.

\begin{figure*}
    \centering
    \begin{subfigure}[t]{0.4\textwidth}
        \centering
        \includegraphics[width=\textwidth]{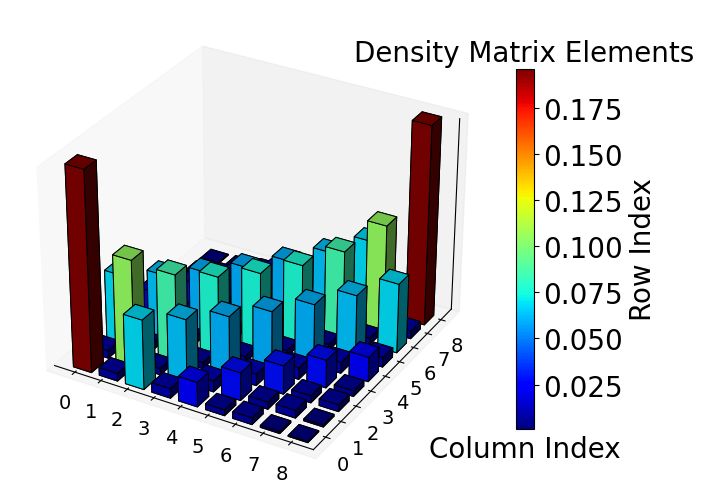}
        \caption{}
    \end{subfigure}%
    ~ 
    \begin{subfigure}[t]{0.35\textwidth}
        \centering
        \includegraphics[width=\textwidth]{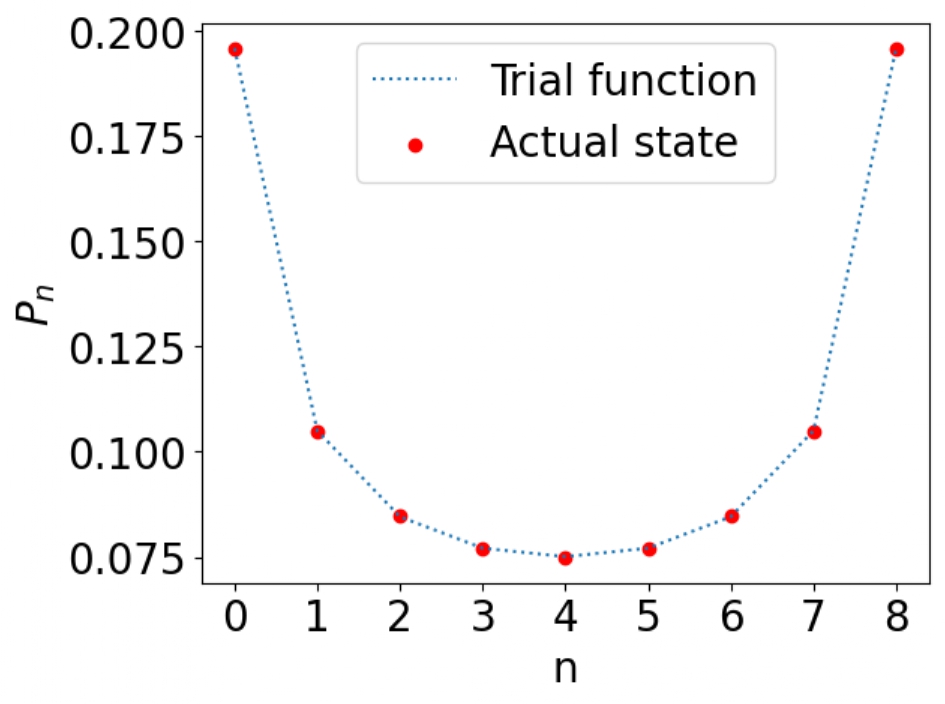}
        \caption{}
    \end{subfigure}
    \caption{For a battery of $N = 8$ qubits under strong driving ($F/g = 10$): (a) Absolute values of the steady-state density matrix elements of the battery in the symmetric Dicke basis. (b) Populations fitted to the trial function given by Eq.~\eqref{eq:pnsd} using the optimal parameters $\alpha = 0.00104$, $\beta = 0.0107$, $\mu = -0.637$, and $\zeta = 0.0345$. With these optimal parameters, the fidelity value of the trial state given by Eq.~\eqref{eq:pnsd} is 0.9.} The other parameters are the same as in Fig.~\ref{fig:energy and ergotropy for F = 0.5}.
    \label{fig:Density matrix elements for F = 10}
\end{figure*}

\section{\label{sec:Results} Results}

In this section, we present the key findings of our study, focusing on how the figures of merit of the battery vary with the number of qubits ($N$). We obtain our results by numerical computation using the  QuTiP~\cite{JOHANSSON20121760} package in Python.

In Fig.~\ref{fig:energy and ergotropy for F = 0.5}, we show how the total energy and ergotropy of the battery in the steady state vary with the number of qubits ($N$) under the intermediate driving strength ($F \sim g$). It is evident that both quantities scale linearly with $N \gg 1$. Similar linear trends can be observed for strong ($F \gg g$) and weak ($F \ll g$) driving regimes. Notably, in all driving regimes, both the total energy and the ergotropy scale linearly with $N$ with the same slope.

A more insightful quantity to examine is the ergotropy-to-energy ratio of the battery, as it directly reflects the quality of charging. The steady-state values of this ratio for different driving strengths $F$ relative to the coupling constant $g$ are shown in Fig.~\ref{fig:Steady-state ergotropy relative to the total energy}. We reiterate that calculating the steady state of the charger-battery system with charger dephasing requires a numerical simulation of the dynamics [Eq.~\eqref{maste eq}] since the steady state is not unique for resonant driving $\omega_\mathrm{B} = \omega_{\mathrm{C}} = \omega_{\mathrm{d}}$. Note that the steady state depends only on the ratio $F/g$~\cite{shastri2025dephasing}. These plots clearly demonstrate that $\mathcal{E}_{\mathrm{B}}/E_{\mathrm{B}}$ increases with the number of qubits across all driving regimes. For small $N$, this growth is most pronounced under weak driving, followed by intermediate and then strong driving. In the large-$N$ limit, the ratio approaches unity as
\begin{equation}
\frac{\mathcal{E}_{\mathrm{B}}}{E_{\mathrm{B}}} \sim 1 - \frac{a}{N} \label{eq:AsymptoticFreedomExpression},
\end{equation}
where $a>0$ is a constant that quantifies the leading-order finite-size ergotropy deficit (see Appendix~\ref{app:A} for an analytical proof of this behavior). A larger value of $a$ indicates that a greater fraction of energy is initially stored in non-work-extractable form, resulting in a slower approach of the ergotropy-to-energy ratio toward unity with increasing $N$. From the fits, we obtain $a \approx 1.38$ at $F/g=10$ in the strong driving regime, $a \approx 1.10$ at $F/g=0.5$ in the intermediate regime, and $a \approx 1.02$ at $F/g=0.2$ in the weak regime. We note that numerical simulations for large $N$ and weak driving is challenging due to two reasons. First, as shown in \cite{shastri2025dephasing}, the optimal charging timescale $\tau^\star$ (for $N=1$) in the weak driving scales as $g \tau^* \sim (F/g)^{-2}$. Moreover, as we discuss below, this charging timescale also has a steep scaling $\tau^\star \sim N^{k}$ (with $k>1$) as a function of the number of batteries. Nonetheless, our numerical calculations clearly indicate that the ergotropy-to-energy ratio in the steady state in this set-up with a dephased charger tends to $1$ in the large $N$ limit --- i.e., we can have asymptotic freedom even in the presence of dephasing \cite{Andolina2019,PhysRevA.111.022222}, where the battery state remains mixed (see Appendix~\ref{app:A} for plots of purity of the steady-state that demonstrate mixedness). This is our central result. Moreover, the ergotropy-to-energy ratio reaches unity most rapidly in the weak driving regime, whereas the strong driving regime exhibits the slowest convergence to asymptotic freedom.

\begin{figure*}
    \centering
    \begin{subfigure}[t]{0.4\textwidth}
        \centering
        \includegraphics[width=\textwidth]{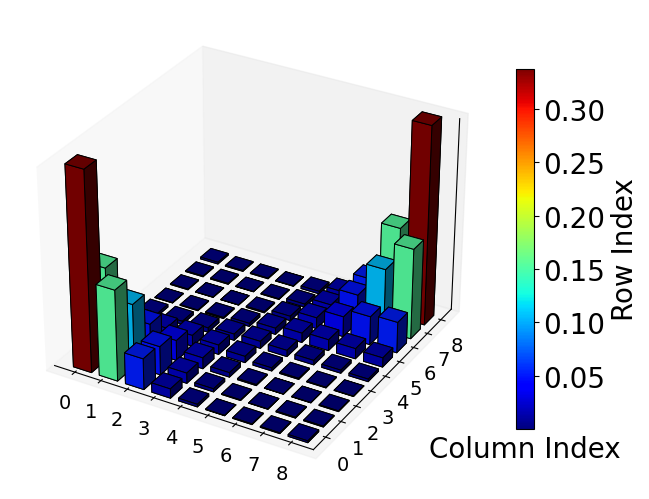}
        \caption{}
    \end{subfigure}%
    ~ 
    \begin{subfigure}[t]{0.4\textwidth}
        \centering
        \includegraphics[width=\textwidth]{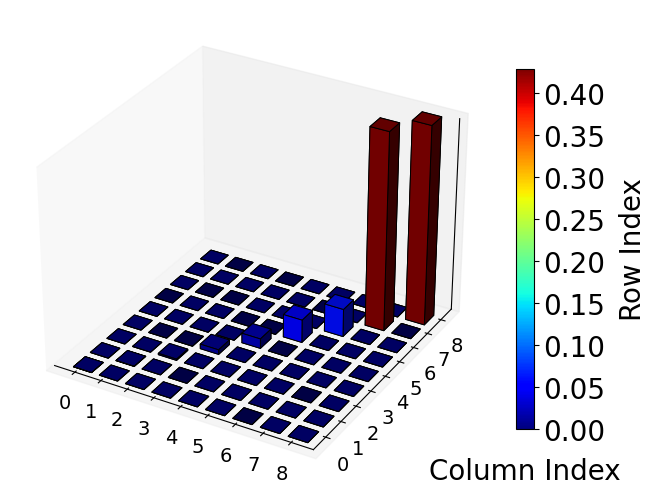}
        \caption{}
    \end{subfigure}
    \caption{For $N = 8$ qubits in the battery under intermediate driving ($F/g = 0.5$) (a) Absolute values of the steady-state density matrix elements of the battery in the symmetric Dicke basis. This state can be well approximated by the trial state given by Eqs.~\eqref{eq:rhoid}--\eqref{eq:psi2} with a fidelity of 0.923, using the optimal parameters $\xi_1 = 0.054 + 2.603\,i$ and $\xi_2 = 0.047 - 1.960\,i$. (b) Absolute values of the corresponding passive-state density matrix elements of the battery. The other parameters are the same as in Fig.~\ref{fig:energy and ergotropy for F = 0.5}.
    }
    \label{fig:Density matrix elements for F = 0.5}
\end{figure*}

To understand the reason behind the emergence of asymptotic freedom as well as its dependence on the charger driving $F/g$, we proceed to examine the steady state of the quantum battery. Given the permutation invariant structure of the total Hamiltonian Eq.~\eqref{eq:Hamiltonian} and the dynamics Eq.~\eqref{maste eq}, the density matrix of the battery is best expressed using the symmetric Dicke basis $\lvert J, m \rangle$, where $J = N/2$ and $m \in \{-J, -J+1, \dots, J-1, J\}$. The Dicke states are eigenstates of the collective spin operators $\hat{J}^2_{\mathrm{B}} = \sum_{\alpha=x,y,z}(\hat{J}^{\alpha}_{\mathrm{B}})^2$ (eigenvalue $J(J + 1)$) and $\hat{J}^Z_{\mathrm{B}}$ (eigenvalue $m$), corresponding to the battery, which can be defined as
\begin{align}
\hat{J}^\alpha_{\mathrm{B}} = \sum_{j=1}^N \hat{\sigma}_{\mathrm{B}, j}^{\alpha}, \qquad \alpha = \{x,y,z \}. \label{eq:collopsdefn}
\end{align}
This formulation allows one to restrict the dynamics to the symmetric Dicke subspace, which has dimension $N+1$.

Focusing first on the strong driving regime, the tomogram of the steady-state density matrix in the Dicke basis for $N = 8$ is shown in Fig.~\ref{fig:Density matrix elements for F = 10}(a). The state exhibits its largest populations in the two extremal Dicke levels, $\lvert N/2, N/2 \rangle$ and $\lvert N/2, -N/2 \rangle$, together with a nearly flat baseline in the central region. To quantitatively capture this characteristic ``edge-enhanced'' and symmetric structure, we consider the following trial density matrix: 
\begin{equation}
  \hat{\rho}_{\mathrm{trial}}^{\mathrm{sd}} = \sum_{n=0}^N\, p_n^{\mathrm{trial}}\, \lvert N/2,\, n-N/2\rangle \langle N/2,\, n-N/2 \rvert,
\end{equation}
where $n$ is the Dicke level index and $p_n^{\mathrm{trial}}$ is the trial population function with a symmetric Bose--Einstein-like functional form with a constant offset,
\begin{equation}
p_n^{\mathrm{trial}} = \zeta + \frac{\alpha}{e^{\beta(n-\mu)} - 1} + \frac{\alpha}{e^{\beta[(N-n)-\mu]} - 1}.\label{eq:pnsd}
\end{equation}
Here, $\alpha$ sets the amplitude of the edge enhancement, $\beta$ plays the role of an effective inverse temperature controlling the decay from the edges toward the center, $\mu$ is a shift parameter determining the inward displacement of the enhancement peaks, and $\zeta$ denotes the baseline population in the central Dicke levels. This functional form reproduces the numerically obtained populations with high accuracy and provides a physically intuitive interpretation [see Fig.~\ref{fig:Density matrix elements for F = 10}(b)]: the profile resembles that of two symmetric bosonic modes, each feeding population from opposite ends of the Dicke ladder, while the flat baseline originates from uniform mixing in the central states induced by the combined effects of strong driving and dephasing. The trial density matrix with the optimal parameters yields high fidelity, where the fidelity between two density matrices $\rho$ and $\sigma$ is defined as
$F(\rho, \sigma)=(\operatorname{tr} \sqrt{\sqrt{\rho} \sigma \sqrt{\rho}})^2$ \cite{JozsaJMO1994}.

In Fig.~\ref{fig:Density matrix elements for F = 0.5}, we present a tomogram of the steady-state density matrix under intermediate driving. We observe that the two extreme Dicke levels, $\lvert N/2, N/2 \rangle$ and $\lvert N/2, -N/2 \rangle$, carry the dominant populations—more pronounced than in the strong driving case—while the intermediate Dicke levels are almost unpopulated. This state can be well approximated by the statistical mixture:
\begin{equation}
\hat{\rho}_{\text{trial}}^{\mr{id}} = \frac{1}{2} \left( \hat{\rho}_1^{\mr{id}} + \hat{\rho}_2^{\mr{id}} \right),\label{eq:rhoid}
\end{equation}
where $\hat{\rho}_1^{\mr{id}} = |\psi_1\rangle \langle \psi_1|$, $\hat{\rho}_2^{\mr{id}} = |\psi_2\rangle \langle \psi_2|$, and
$|\psi_1\rangle$ and $|\psi_2\rangle$ are spin-coherent states generated from the extremal Dicke states given by
\begin{align}
|\psi_1\rangle &= \frac{e^{\xi_1 \hat{S}_-} \lvert N/2, ~N/2 \rangle}
{\big\| e^{\xi_1 \hat{S}_-} \lvert N/2, N/2 \rangle \big\|},\label{eq:psi1}\\
|\psi_2\rangle &= \frac{e^{\xi_2 \hat{S}_+} \lvert N/2, -N/2 \rangle}
{\big\| e^{\xi_2 \hat{S}_+} \lvert N/2, -N/2 \rangle \big\|},\label{eq:psi2}
\end{align}
with $\xi_1, \xi_2 \in \mathbb{C}$ being complex parameters.

To understand the observed increase in the ergotropy-to-energy ratio 
$\left(\mathcal{E}_{\mathrm{B}}/E_{\mathrm{B}}\right)$ with the number of qubits $N$, 
we examine the passive state of the battery, obtained by minimizing the energy over all unitary operations within its $2^N$-dimensional Hilbert space. In this passive state, most of the population resides in the ground and first excited levels. For presentation purposes, Fig.~\ref{fig:Density matrix elements for F = 0.5}(b) shows the passive state in the energy basis to illustrate this population concentration; the ergotropy itself is, however, computed in the full $2^N$-dimensional Hilbert space. Since we are studying the ratio $\mathcal{E}_{\mathrm{B}}/E_{\mathrm{B}}$, a more relevant quantity to consider is the ratio of the energy gap between the first excited state and the ground state ($\Delta_{\mathrm{g}}$) to the total energy of the battery, denoted by $\Delta_{\mathrm{g}}/E_{\mathrm{B}}$, where $\Delta_{\mathrm{g}} \sim \omega_{\mathrm{B}}$ and $E_{\mathrm{B}} \sim N \omega_{\mathrm{B}}$. This implies that $\Delta_{\mathrm{g}}/E_{\mathrm{B}} \sim 1/N$, which tends to zero in the large-$N$ limit. In fact, as $N \to \infty$, the ground state becomes approximately degenerate, and the ratio $\mathcal{E}_{\mathrm{B}}/E_{\mathrm{B}} \to 1$ even though the state remains mixed. The emergence of approximate ground-state degeneracy in the large-$N$ limit leads to asymptotically maximal ergotropy relative to the total energy, or ``asymptotic freedom,'' despite the non-purity of the state.

In the intermediate driving regime, the two extreme Dicke levels carry the dominant populations, while the intermediate Dicke levels remain almost unpopulated. Consequently, in the passive state of the battery, most of the population is concentrated in the ground and first excited states, leading to a faster increase of the $\mathcal{E}_{\mathrm{B}}/E_{\mathrm{B}}$ ratio with increasing $N$. In contrast, in the strong driving regime, the steady state of the battery has non-negligible populations in the intermediate Dicke levels, which translates into appreciable populations in the higher-energy levels of the passive state. As a result, the $\mathcal{E}_{\mathrm{B}}/E_{\mathrm{B}}$ ratio increases with $N$ considerably more slowly than in the intermediate regime.

\begin{figure*}
    \centering
    \begin{subfigure}[t]{0.30\textwidth}
        \centering
        \includegraphics[width=\textwidth]{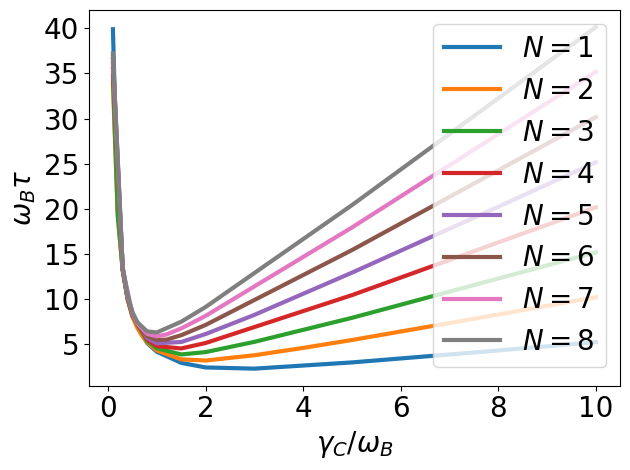}
        \caption{}
    \end{subfigure}%
    ~ 
    \begin{subfigure}[t]{0.30\textwidth}
        \centering
        \includegraphics[width=\textwidth]{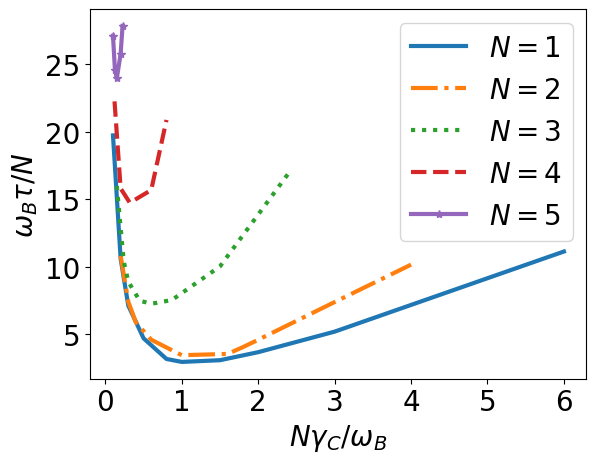}
        \caption{}
    \end{subfigure}
    ~ 
    \begin{subfigure}[t]{0.30\textwidth}
        \centering
        \includegraphics[width=\textwidth]{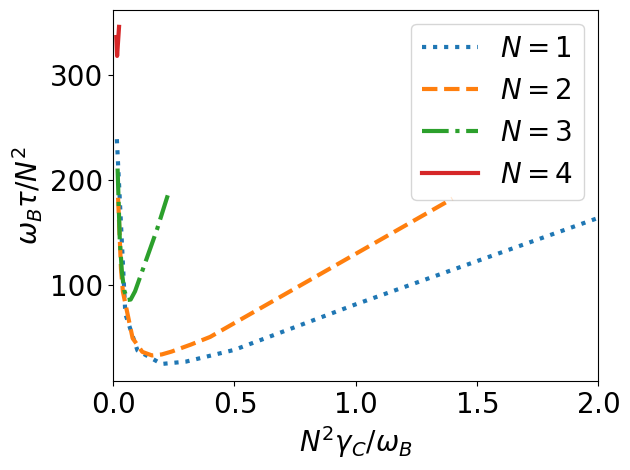}
        \caption{}
    \end{subfigure}
    \caption{Charging time $\tau$ of the battery as a function of the charger dephasing rate $\gamma_{\mathrm{C}}$ for different driving regimes: (a) strong driving $F = 10 \omega_{\mathrm{B}}$ ($F/g = 10$), (b) intermediate driving $F = 0.5 \omega_{\mathrm{B}}$ ($F/g = 0.5$), and (c) weak driving $F = 0.2 \omega_{\mathrm{B}}$ ($F/g = 0.2$). The $x$- and $y$-axis scalings in panels (b) and (c) are adjusted so that results for different $N$ can be displayed within the same region. Other parameters are the same as in Fig.~\ref{fig:energy and ergotropy for F = 0.5}.
    }
    \label{fig:charging time}
\end{figure*}

\begin{figure*}
    \centering
    \begin{subfigure}[t]{0.3\textwidth}
        \centering
        \includegraphics[width=\textwidth]{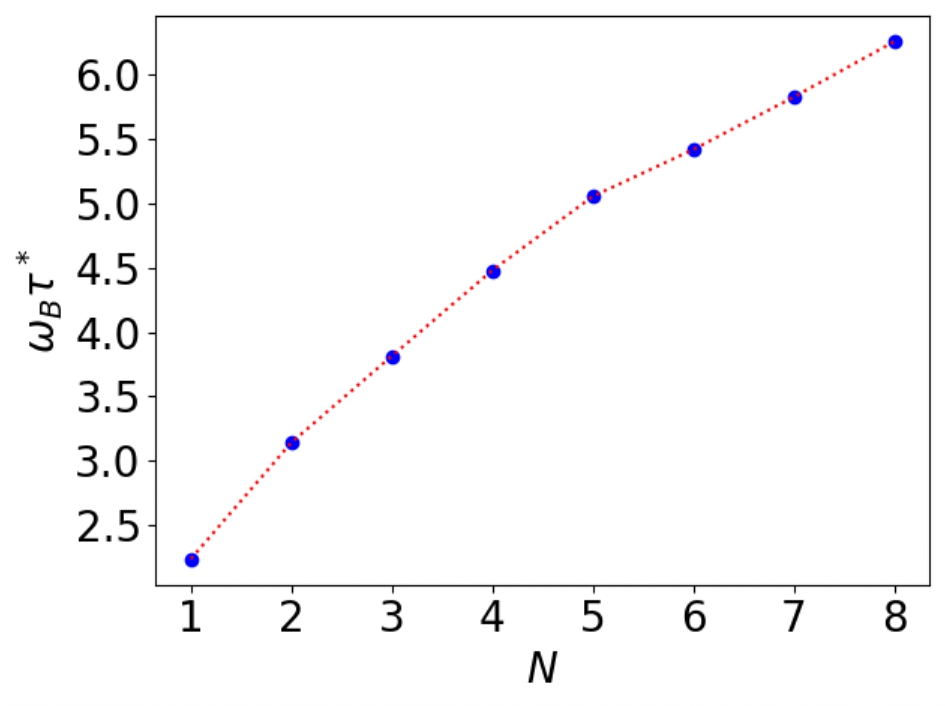}
        \caption{}
    \end{subfigure}%
    ~ 
    \begin{subfigure}[t]{0.3\textwidth}
        \centering
        \includegraphics[width=\textwidth]{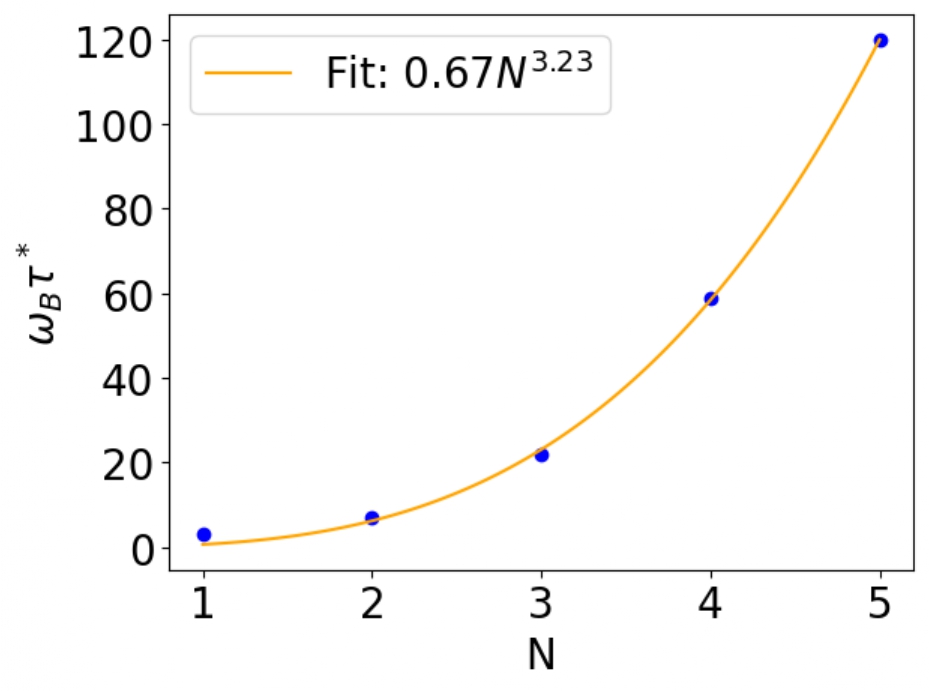}
        \caption{}
        \end{subfigure}
        ~ 
    \begin{subfigure}[t]{0.3\textwidth}
        \centering
        \includegraphics[width=\textwidth]{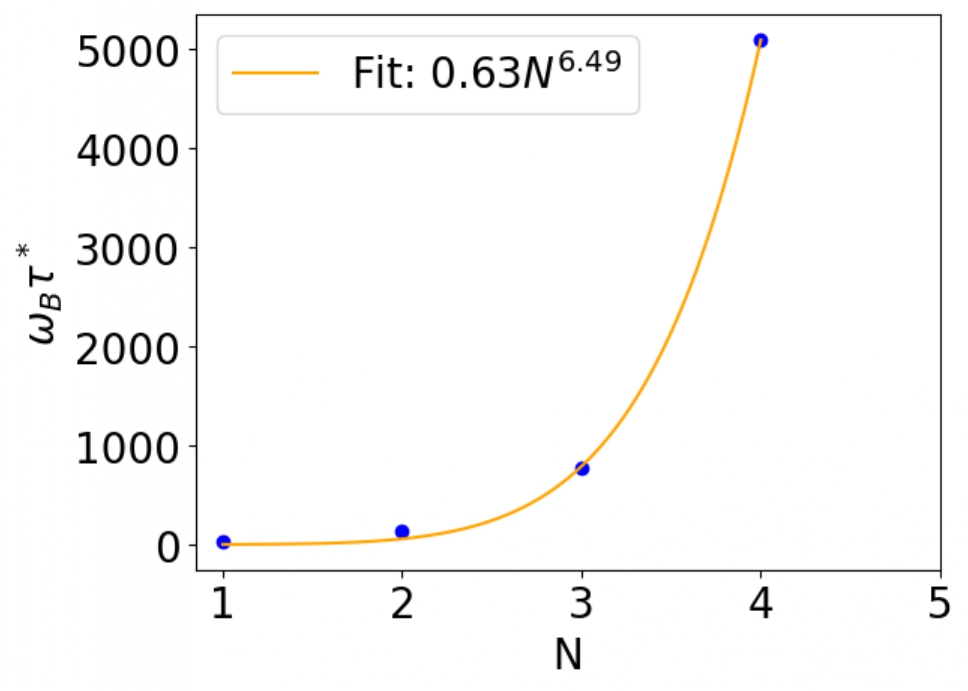}
        \caption{}
        \end{subfigure}
    \caption{Optimal charging time $\tau^*$ of the battery as a function of the number of qubits in the battery for different regimes of the driving strength: (a) strong driving $F = 10 \omega_{\mathrm{B}}$ $(F/g = 10)$, (b) intermediate driving $F = 0.5 \omega_{\mathrm{B}}$ $(F/g = 0.5)$, and (c) weak driving $F = 0.2 \omega_{\mathrm{B}}$ $(F/g = 0.2)$. The other parameters are the same as in Fig.~\ref{fig:energy and ergotropy for F = 0.5}.}
    \label{fig:optimal charging time}
\end{figure*}

In Fig.~\ref{fig:charging time}, we show the charging time $\tau$ of the battery as a function of the charger's dephasing rate for different driving strengths. The corresponding optimal charging time, $\tau^*$, is defined as the minimum value of $\tau$ at the optimal dephasing rate $\gamma_\mathrm{C}^*$ of the charger~\cite{shastri2025dephasing}. Figure~\ref{fig:optimal charging time} presents $\tau^*$ as a function of the number of qubits $N$. In the strong-driving regime, the optimal charging time grows approximately linearly with $N \gg 1$, whereas in the intermediate and weak driving regimes, it follows power-law scalings of $\tau^* \approx 0.67\,N^{3.23}$ at $F/g = 0.5$ and $\tau^* \approx 0.63\,N^{6.49}$ at $F/g = 0.2$, respectively.

Strong driving can simultaneously excite multiple qubits, allowing the system to access intermediate Dicke states without strictly climbing the ladder one step at a time. These multi-excitation pathways open several parallel channels for charging, enabling faster energy transfer and yielding an almost linear scaling of the charging time with $N$. In contrast, weak and intermediate driving cannot induce multi-qubit excitations, as the drive is too weak to activate fast multi-photon or multi-qubit channels. In these regimes, sequential collective excitations dominate: the system must traverse a long sequence of single-step transitions, each with a limited rate, resulting in a slower charging process and superlinear growth of the charging time with $N$. 
Moreover, as we saw earlier, in such cases, the steady state is dominated by the two extreme Dicke states $\ket{J, \pm J}$ with negligible population in the intermediate levels.


\section{\label{sec:Conclusion} Conclusion}

Quantum batteries are typically studied in closed-system settings, which are idealized and often exhibit oscillatory charging dynamics. In contrast, open-system setups can enable stable charging, where an optimal pure dephasing on a driven quantum charger can significantly accelerate the charging process. However, such setups often suffer from low charging quality, as a large fraction of the stored energy remains locked and cannot be extracted as work due to the reduction of the battery-state purity caused by decoherence.

In this work, we have investigated how an ``asymptotic freedom''-like behavior—where all deposited energy can be extracted as work, resulting in the ergotropy-to-energy ratio approaching unity—can emerge in the steady-state charging of a quantum battery. We considered a collective charging setup in which \( N \) qubits are coupled to a driven qubit charger in a star configuration, with controlled pure dephasing acting on the charger. We analyzed the steady-state ergotropy-to-energy ratio across different driving regimes and found that it increases with the number of qubits and approaches unity asymptotically as $1-O(1/N)$ in all cases. The increase is most pronounced under weak driving, followed by intermediate and strong driving. In the large-$N$ limit, the emergence of approximate ground-state degeneracy leads to asymptotically maximal ergotropy relative to the total energy—i.e., asymptotic freedom—even though the battery state remains mixed.

We also investigated the charging time $\tau$, which quantifies how quickly the battery approaches its steady-state energy. The optimal charging time $\tau^*$, obtained at the charger’s optimal dephasing rate $\gamma_{\mathrm{C}}^*$, exhibits distinct scaling behaviors with the number of qubits $N$: $\tau^* \sim N$ under strong driving, and $\tau^* \sim N^b$ with $b>1$ in intermediate and weak driving. Under strong driving, multi-qubit excitations and parallel pathways via intermediate Dicke states enable fast, nearly linear charging. By contrast, weak and intermediate driving rely on sequential collective excitations through the extremal Dicke states $|J,\pm J\rangle$, resulting in superlinear growth of $\tau^*$.

Therefore, if the goal is high-quality charging, weak or intermediate driving is better, despite the superlinear growth of $\tau^*$. For fast charging, strong driving is preferable for large $N$, even though the ergotropy-to-energy ratio is lower. This highlights a fundamental speed–quality trade-off in collective systems with dephased charging.

Finally, we note an interesting qualitative change in the steady-state structure of the battery when moving from intermediate to strong driving [see Figs.~\ref{fig:Density matrix elements for F = 0.5}(a) and \ref{fig:Density matrix elements for F = 10}(a)]. In the strong driving case, the intermediate Dicke levels acquire appreciable populations, whereas under intermediate driving the steady state is dominated by the two extreme Dicke levels with negligible intermediate-level population. Such a restructuring of the steady state may be indicative of a nonequilibrium phase transition in the large-$N$ limit. A more rigorous characterization of this transition, and its relation to the charging dynamics, remains an interesting direction for future research.

\section*{Data availability statement}
All data generated during this study can be reproduced using the described methodology.

\begin{acknowledgments}
G.W. was supported by the National Natural Science Foundation of China (Grants No.~12375039 and No.~11975199). PV acknowledges support from MATRICS Grant No.~MTR/2023/000900 from the Anusandhan National Research Foundation, Government of India.
\end{acknowledgments}
\begin{figure*}
    \centering
    \begin{subfigure}[t]{0.3\textwidth}
        \centering
        \includegraphics[width=\textwidth]{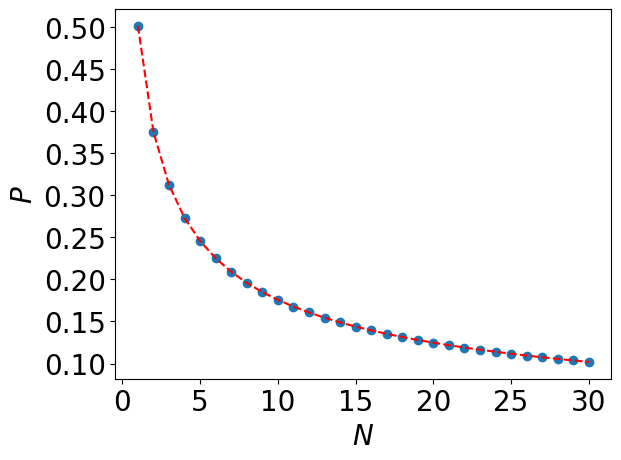}
        \caption{}
    \end{subfigure}%
    ~ 
    \begin{subfigure}[t]{0.3\textwidth}
        \centering
        \includegraphics[width=\textwidth]{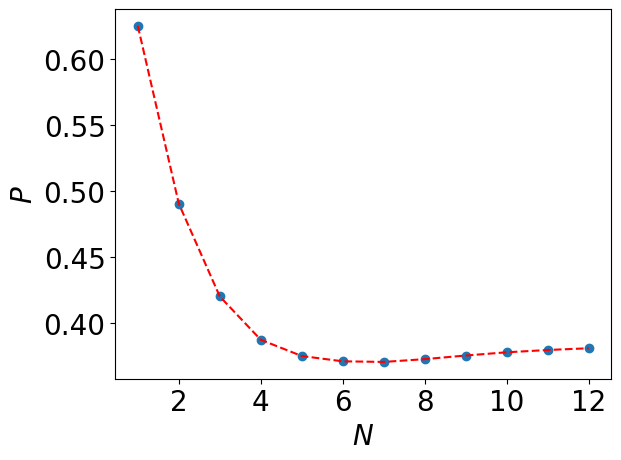}
        \caption{}
        \end{subfigure}
        ~ 
    \begin{subfigure}[t]{0.3\textwidth}
        \centering
        \includegraphics[width=\textwidth]{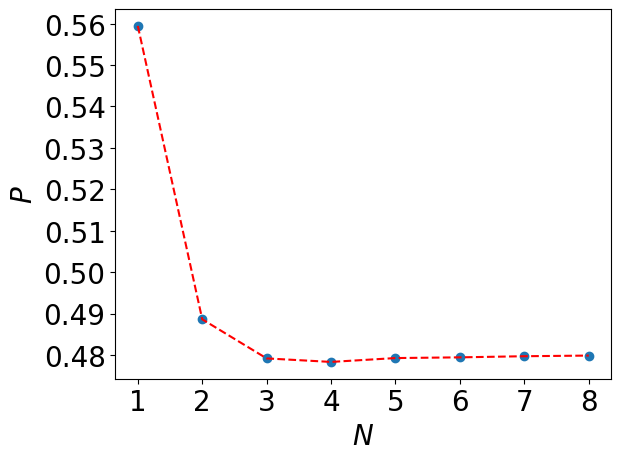}
        \caption{}
        \end{subfigure}
    \caption{ Variation of purity $P \equiv \operatorname{Tr}[{\hat{\rho}_{\mathrm{B}}^2}]$ of the state of the battery in the steady state as a function of the number of qubits $N$ in the battery for different regimes of the driving strength: (a) strong driving $F = 10 \omega_{\mathrm{B}}$ $(F/g = 10)$, (b) intermediate driving $F = 0.5 \omega_{\mathrm{B}}$ $(F/g = 0.5)$, and (c) weak driving $F = 0.2 \omega_{\mathrm{B}}$ $(F/g = 0.2)$. The other parameters are $\omega_{\mathrm{B}} = \omega_{\mathrm{C}} = \omega_{\mathrm{d}} = 1$ and $g = 1.0\,\omega_{\mathrm{B}}$.}
    \label{fig:purity}
\end{figure*}
\begin{figure*}
    \centering
    \includegraphics[width=0.93 \linewidth]{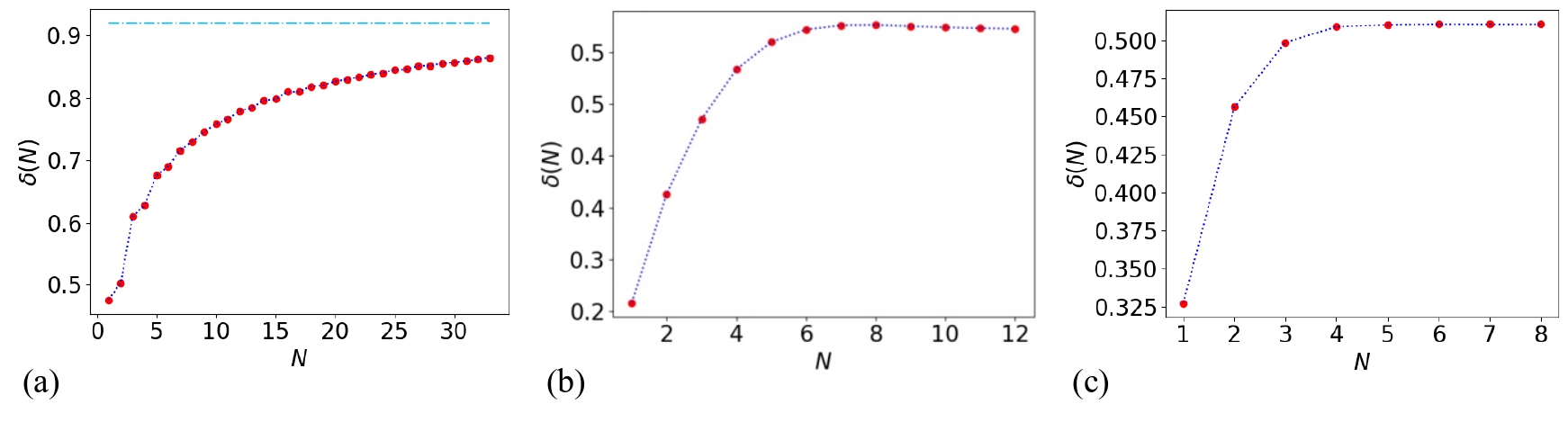}
    \caption{Variation of the population $\delta(N)$ outside the ground state for the passive state of the battery corresponding to the steady state as a function of the number of qubits $N$, for different regimes of the driving strength: (a) strong driving $F = 10 \omega_{\mathrm{B}}$ $(F/g = 10)$ (red dots) and the horizontal dash-dotted line (deepskyblue color) shows the asymptotic value of $\delta(N)$ in the limit of $N \to \infty$, (b) intermediate driving $F = 0.5 \omega_{\mathrm{B}}$ $(F/g = 0.5)$, and (c) weak driving $F = 0.2 \omega_{\mathrm{B}}$ $(F/g = 0.2)$. The other parameters are $\omega_{\mathrm{B}} = \omega_{\mathrm{C}} = \omega_{\mathrm{d}} = 1$ and $g = 1.0\,\omega_{\mathrm{B}}$.}
    \label{fig:DeltavsN}
\end{figure*}

\appendix
\section{\label{app:A} Asymptotic scaling behavior of $\mathcal{E}_{\mathrm{B}}/E_{\mathrm{B}}$}

The spectral decompositions of the battery state $\hat{\rho}_\mathrm{B}$ and its Hamiltonian $\hat{H}_\mathrm{B}$ are given in Eq.~(\ref{Spectal}). The corresponding passive state is
\begin{equation}\label{passive_state}
\hat{\rho}_{\mathrm{B}}^{\downarrow}=\sum_i \eta_i^{\downarrow}\, \ket{\varepsilon_i^{\uparrow}}\!\bra{\varepsilon_i^{\uparrow}},
\end{equation}
whose energy is
\begin{equation}
\operatorname{Tr}\!\left[\hat{H}_{\mathrm{B}}\, \hat{\rho}_{\mathrm{B}}^{\downarrow}\right]
=\sum_i \varepsilon_i^{\uparrow}\, \eta_i^{\downarrow}. \label{eq:passiverg}
\end{equation}
If $\hat{\rho}_{\mathrm{B}}$ is pure, the minimum is attained for $\eta_0^{\downarrow}=1$ and $\eta_{i\neq 0}^{\downarrow}=0$, yielding 
$\operatorname{Tr}\!\left[\hat{H}_{\mathrm{B}}\, \hat{\rho}_{\mathrm{B}}^{\downarrow}\right]=0$.
For a mixed state, necessarily $\eta_0^{\downarrow}<1$ and some $\eta_{i\neq 0}^{\downarrow}>0$. In our setup, the steady state $\hat{\rho}_{\mathrm{B}}$ is mixed, even as $N\to\infty$ (see Fig.~\ref{fig:purity}). Thus, the population outside the lowest energy eigenstate in the passive state $\hat{\rho}_{\mathrm{B}}^{\downarrow}$ for the set-up with $N$-qubits,
\begin{align}
\delta(N) \equiv 1- \eta_0^{\downarrow},
\end{align}
always satisfies $0< \delta(N) \leq 1$. Moreover, as we show in Fig.~\ref{fig:DeltavsN}, $\delta(N)$ tends to a nonzero value as asymptotically, i.e., $\delta \equiv \lim_{N\to\infty} \delta(N)>0$ (for all driving strength regimes). 

Focusing on the passive energy given by Eq.~\eqref{eq:passiverg}, we note that in general the sum over $i$ extends from $0$ to $2^N-1$ owing to the dimensionality of the computational basis for the $N$-qubit battery system being $2^N$. However, in our case, since the equation of motion in Eq.~\ref{maste eq} and the initial state are permutation symmetric, the density matrix has nonzero eigenvalues only in the permutation-invariant subspace $\mathcal{S}$ generated by the Dicke basis which has the dimension $\operatorname{dim} \mathcal{S}= N+1$. Thus we have that
\begin{align}
    \operatorname{Tr}\left[\hat{H}_B \hat{\rho}_B^{\downarrow}\right]
&=\sum_{i=0}^{N} \varepsilon_i^{\uparrow} \eta_i^{\downarrow}. \label{eq:permsymmergpass}
\end{align}
Now, the spectrum of the Hamiltonian $(\hat{H}_{\mathrm{B}})$ is given as $\bar{\varepsilon}_n=n \omega_{\mathrm{B}}$, where each eigenvalue has multiplicity ${ }^N C_n$. Due to their multiplicity, to calculate the sum over the first $N+1$ eigenvalues [a subspace of dimension $\operatorname{dim} \mathcal{S}$ in the sum of Eq.~(\ref{eq:permsymmergpass})], it is sufficient to consider the first two distinct eigenvalues $\left(\bar{\varepsilon}_0=0\right.$ and $\bar{\varepsilon}_1=\omega_{\mathrm{B}}$), where $ { }^N C_0+{ }^N C_1 = N+1$. Thus, we have
\begin{align}
    \operatorname{Tr}\left[\hat{H}_B \hat{\rho}_B^{\downarrow}\right]
&= 0  \times \eta_0^{\downarrow}  + \omega_\mathrm{B}\sum_{i=1}^{N}\eta_i^{\downarrow} \nonumber \\
&= \omega_\mathrm{B} (1-\eta_0^{\downarrow}) = \omega_\mathrm{B} \delta (N). 
\end{align}
Using the definition of the ergotropy, $\mathcal{E}_\mathrm{B} = E_\mathrm{B} - \operatorname{Tr}\left[\hat{H}_B \hat{\rho}_B^{\downarrow}\right]$, and the fact that $E_{\mathrm{B}}= N\,\omega_{\mathrm{B}}/2$ in our scheme (see Fig.~\ref{fig:energy and ergotropy for F = 0.5}), we obtain
\begin{align}
    \frac{\mathcal{E}_\mathrm{B}}{E_\mathrm{B}} = 1 - \frac{2\delta(N)}{N}.\label{eq:ratioequality}
\end{align}
Combined with the behaviour of $\delta(N)$ shown in Fig.~\ref{fig:DeltavsN}, this leads to the asymptotic result (for $N \rightarrow \infty$) 
\begin{align}
    \frac{\mathcal{E}_\mathrm{B}}{E_\mathrm{B}} \sim 1 - \frac{2\delta}{N}, 
\end{align}
thereby proving the claim in Eq.~\eqref{eq:AsymptoticFreedomExpression} with $a = 2\delta$. 
Finally, we also note that the above proof does not depend on the type of the charger (cavity mode or qubit) once we trace out the charger and end up with the state of the batteries.

\bibliography{mybib}

\end{document}